# Topological Magnetoelectric Switching on the Atomic Scale

Igor Altfeder[1*], Robert C. Walko[1], Seng Huat Lee[2,3] and Zhiqiang Mao[2,3]

[1]Department of Physics, Ohio State University, Columbus, OH 43210, USA

[2]Department of Physics, Pennsylvania State University, University Park, PA 16802, USA

[3]2D Crystal Consortium, Materials Research Institute, Pennsylvania State University, University Park, PA 16802, USA

*Igor.Altfeder@gmail.com

**Abstract**

Intrinsic magnetic topological insulators from $MnBi_2Te_4$ family promise realization of axion insulator, a theoretically predicted state of matter capable to mutually convert electric and magnetic signals, and whose properties are reminiscent of cosmological axion dark matter. Here we report the first experimental realization of room-temperature axionic single-spin switch in topological metamaterial $MnBi_2Te_3$. The metamaterial was in situ created in the scanning tunneling microscopy (STM) experiment by removing Te atomic layer from $MnBi_2Te_4$ surface using STM tip. Room temperature STM study of $MnBi_2Te_3$ revealed atomic scale variations of exchange gap and formation of nanoscale spin bubbles pinned at subsurface defects. We found that individual spin states in $MnBi_2Te_3$ can be reversibly switched using local electric field of STM tip with magnetoelectric response comparable to the theoretically predicted response of axion insulator. The observed topological surface magnetism develops significantly above bulk Néel temperature.

**Introduction**

Magnetically gapped topological Dirac electrons [1-20] promise realization of axion insulator, a theoretically predicted state of matter capable to convert electric field into magnetic polarization [9, 10, 20], carry quantum anomalous Hall transport [7-11], and whose properties are reminiscent of cosmological axion dark matter [6]. In magnetically doped topological insulators (TIs) [7, 8, 16] the magnetic gapping is local and typically disordered. Currently, the search for axion insulator shifts towards intrinsic magnetic TI MnBi$_2$Te$_4$ [1-5, 12-15], where Mn atoms form a complete atomic layer (see Fig. 1a) and topological states become fully gapped with E$_g$ ~ 0.1 eV [1, 5, 15]. MnBi$_2$Te$_4$ crystallizes in a rhombohedral structure with the space group $R\bar{3}m$ [3], built of the stacking of Te-Bi-Te-Mn-Te-Bi-Te septuple layers (SLs) along the *c*-axis as shown in Fig. 1a. The antiferromagnetism (AFM) of this material with bulk T$_N$=25 K [14] is produced by Mn sublattice, while its nontrivial surface state is formed by inverted Bi and Te p$_z$ bands at the Γ point due to strong spin-orbital coupling. According to recent experiments, even at T>>T$_N$ MnBi$_2$Te$_4$ surface [15] shows signature of strong two-dimensional (2D) ferromagnetic (FM) fluctuations, indicating that its gapped topological surface states (TSS) can amplify magnetic interactions [17-19] and its further exploration may lead to room temperature (RT) axion insulator. A possible path towards realization of this goal is suggested in Fig. 1b. The complex structure of SL allows designing topological magnetic metamaterials terminated by other than van der Waals (vdW) Te plane. In the proposed in Fig. 1b MnBi$_2$Te$_3$/MnBi$_2$Te$_4$ heterostructure the terminating layer of Te atoms is removed using scanning tunneling microscope (STM) as bottom-up nanofabrication tool [21]. It is anticipated that the asymmetric structure of the outmost *sextuple* layer will result in modification of gapped TSS and enhancement of interactions. Study of asymmetric layers may also shed light on physics of magnetic defects [2] and canted spins [15] on the atomic scale. Due to its exclusive surface sensitivity and resolution, STM represents the most relevant tool for exploring these effects [4, 16, 21-25]. Here we report the first realization of RT axionic spin switch in topological quantum metamaterial using local electric field of STM tip.

## Results

The RT STM image of cleaved Te-terminated $MnBi_2Te_4$ is shown in Fig. 2a. The surface contains atomically-flat-bottom nanocorrals (voids) of 2-4 nm lateral size. The 2D Fourier transform of this image, shown in the inset of Fig. 2a, reveals 4.3 Å unit cell vectors as anticipated for $MnBi_2Te_4$ [4]. The cross-sections of STM images indicated that the depth of nanocorrals is 6.1 Å, as illustrated in Fig. 2b, and their bottoms correspond to Mn layer [15]. The subsequent removal of the terminating Te atomic layer was achieved using empirically established rapid STM lithography method by increasing the tunneling current and inducing sub-SL cleavage as described in Methods and in the Supplement. The inset of Fig. 2c shows the 1700×1700 Å$^2$ STM image obtained after removing Te atomic layer. The terminating layer was removed at ≈ 500×500 Å$^2$ part of the surface, and the exposed area is indicated by a dotted square. STM image of the exposed surface area is shown in Fig. 2c. The image sizes are 400×400 Å$^2$. The nanocorrals with typical lateral sizes as in Fig. 2a still can be observed, although their depth has decreased. The STM cross-sections of nanocorrals (Fig. 2d) have shown that their depth decreased by ≈35%, as anticipated for Te layer removed [15] (Bi-Te vs. Te-Bi-Te). The $MnBi_2Te_3$/$MnBi_2Te_4$ heterostructure represents only one example of metamaterials that can be in-situ fabricated using STM. We have conducted more STM lithography experiments (see Supplement Fig. 3Sup) and found that multiple non-vdW terminated surfaces can be obtained.

Fig. 3a shows high-resolution STM image of the exposed Bi-terminated surface area. We noticed that some of the atoms look brighter than others. These are substitution defects in the Bi atomic plane, often associated with antisites [22]. For comparison, the inset of Fig. 3a shows STM image of pristine Te-terminated surface with triplet electronic defects that were previously reported in MnBi$_2$Te$_4$ [4, 22] and magnetically doped TIs [16] and are possibly related to antisites located in the Bi atomic plane. Their density is comparable to the density of substitution defects directly observable in Fig. 3a. The example of STM imaging atomic lattices both outside and inside nanocorrals (Bi- vs. Mn-termination) is shown in Fig. 3b.

Besides substitution defects, STM images of Bi-terminated heterostructure revealed elliptic "bubbles" of ≈2-4 nm lateral size, whose examples are shown in Figs. 3c and 3d. The STM cross-section in Fig. 3e allows comparing the geometries of a bubble and a patch of substitution defects located in its vicinity. While the substitution defects produce sharp steps on the cross-section (left side), the profile of the bubble (right side) is smooth and stepless. The peak STM signal contrast in the center of the bubble ≈2 Å indicates strong local density of states (DOS) variations. These bubbles were found to be scattered all over the surface and in some instances form clusters or chains (see lower inset in Fig. 4a and Fig. 4Sup). To understand the origin of these structures, tunneling current-voltage (I-V) measurements were performed at various locations inside and around the bubbles. The typical I-V characteristic is shown in the upper inset of Fig. 4a. The curves are acquired with a tunnel bias (sample-to-tip bias) sweep from -0.5V to +0.5V and then reversed back, with tip-sample distance held constant. The hysteresis loop can be clearly observed at positive bias; its direction is indicated by dotted arrows. The main panel in Fig. 4a shows the same I-V curves using linear-logarithmic vertical scale $arsinh(I)$. This earlier established [23, 24] linearization procedure eliminates the exponential increase of tunneling current at elevated bias typically caused by change of the shape of the tunneling barrier. After linearization, in addition to the hysteresis loop at positive bias, the bandgap $E_g$ can also be observed below the Fermi level (FL). The analysis of observed I-V dependence is presented in Figs. 4b and 4c. In Fig. 4b, applying positive tunneling bias leads to tilting of the tunneling barrier and decreases its effective width, which causes exponential I-V dependence [25], i.e. $ln(I) \propto V$ dependence observed in Fig. 4a below $V_{on} \approx 0.4\,V$. The FL remains pinned at the conduction band minimum (CBM), indicating that below $V_{on}$ the electric field of the tip causes only partial depletion of surface n-type carriers. Above this threshold, as we show in Fig. 4c, a complete depletion of surface n-type carriers takes place under the STM tip and the FL is located inside the gap (metal-to-insulator transition). Now the further increase of the tunneling bias leads to band-bending under the tip, whereas the barrier tilting stops resulting in $I(V) \approx$ const. Tunneling I-V measurements at larger tip-sample distances ($R$) show the decrease of saturation current and the overall shift of hysteresis loop towards higher voltages, which allows to determine the electric field required for metal-to-insulator transition $\partial V_{on}/\partial R \approx 10^9$ V/m (see also Fig. 5Sup), in good agreement with earlier found n-type carrier density $1.3 \times 10^{20}$ cm$^{-3}$ [15]. In Fig. 4c, the exchange gap after the transition is shown enlarged; the gap increase is what causes the hysteresis loop observed on the reverse I-V sweep. Such an increase of $E_g$ may be triggered by interaction of tip-induced electrostatic field with subsurface Mn atomic layer (interaction that was screened before metal-to-insulator transition). To verify the magnetic nature of the hysteresis loops, the I-V characteristics were measured at different distances from the center of the bubble. The locations of I-V measurements are along the dotted line on the image in Fig. 5a. The spatial variations of both hysteresis loop width δ and $E_g$ can be observed in the I-V characteristics shown in Fig. 5d. The curve 1 shows zero δ and zero $E_g$ that corresponds to (pinned) center of the bubble and unlike all other curves does not reveal strong signature of metal-to-insulator transition. Interestingly, upon progression from point 1 to point 5 the exchange gap gradually increases, whereas δ behaves non-monotonically. It jumps from zero to its maximum value at point 2 and then gradually decreases at points 2-5. Except for point 1, we observe complementary behavior of δ and $E_g$, i.e. an increase of one of them is accompanied by a decrease of another. For estimation of $E_g$, the left part of linearized I-V characteristic can be extrapolated to intersection with the horizontal axis. For points 2-5 $E_g$(eV) is 0.15, 0.18, 0.21, 0.23 and δ(eV) is 0.19, 0.15, 0.13, 0.09. Surprisingly, we find that $E_g + \delta \approx const \approx 0.33\ eV$.

For magnetic topological insulator, the exchange gap $E_g = 2|\Delta_M| \propto |M_\perp|$, where $\Delta_M$ is exchange field produced by magnetic sublattice [9, 10, 26] and $M_\perp$ is out-of-plane magnetization [27]. Therefore, our observation of spatially varying exchange gap directly confirms the magnetic nature of the bubbles with spatially modulated $M_\perp$; one of the corresponding spin structures is illustrated in Fig. 5b. Such spatially pinned spin bubbles may develop around subsurface defects with fixed in-plane magnetization, which can explain the observed elliptic shapes (see Supplement P7), and their sizes should be comparable to magnetic domain wall width [28]. Defect induced localized spin canting has been known both in 2D [2] and in 3D materials [29]. For studied here material, the likely defect candidates are Mn vacancies and their clusters [15], and bismuth-vacancy complexes ($Bi_{Mn}V_{Mn}$) in the Mn atomic plane. According to previous theoretical studies [17-19, 30], stabilization of surface magnetism in $MnBi_2Te_3$ is likely to be a combined effect of enhanced exchange [17-19] and enhanced magnetic anisotropy [30, 31] energies due to interaction of magnetic atoms with TSS. In Figs. 5b and 5c we show the field-dependent spin structure of the bubble; the construction takes into account that the bubble size (*w*) can decrease after metal-to-insulator transition of TSS when electric field penetrates towards Mn layer and its magnetic anisotropy energy further increases [32, 33]. The observed variations of δ and $E_g$ inside a bubble can be qualitatively explained using simplified model presented in Fig. 5e. The model approximates radial dependence $M_\perp(r)$ inside a bubble as $M_\perp \propto \sin(\pi r/w)$, which corresponds to uniform rotation of spins from the out-of-plane (θ=0) direction on the periphery $r = w/2$ to in-plane (θ=90°) direction in the center $r = 0$. The dashed and dotted lines correspond to bubble sizes $w_1$=5 nm (zero-field value) and $w_2$=1 nm (in-field value) respectively and show the radial dependences of normalized exchange gap. The zero-field curve predicts a monotonic increase of $E_g$ from the center to periphery of the bubble, as indeed observed in Fig. 5d. The solid line shows the difference between in-field and zero-field $E_g$ curves that quantifies the size of the I-V hysteresis loop. The predicted δ(r) dependence is non-monotonic in excellent agreement with Fig. 5d, which strongly confirms the proposed mechanism of field-induced local spin switching. The experimental point 1 corresponds to $r = 0$ on δ(r) curve; at this location metal-to-insulator transition is not observed and the exchange gap is always zero. For other locations, the observed constant $E_g + \delta$ implies that the bandgap increase mainly occurs at CBM (not symmetrically as in Fig. 4c) indicating that the effective mass ratio of the gapped Dirac cone $m_e/m_h \ll 1$, similar to MnBi$_2$Te$_4$ [1, 10, 34]. Thus, the analysis of our spectroscopic data shows that $E_g$'s from Fig. 5d represent the spatial dependence of $|M_\perp| \propto |\cos\theta|$ in the absence of electric field, whereas $E_g + \delta \approx const$ corresponds to saturated out-of-plane magnetization (θ=0 or π) under strong electric field (see also Fig. 6Sup).

Tunneling measurements inside nanocorrals did not reveal any signature of a gap (see Fig. 7Sup) strongly confirming that $E_g$ is surface related property. From the Fig. 5d data, the peak exchange gap corresponding to $\cos\theta = \pm 1$ can be estimated as $E_G = E_g + \delta = 330\ meV$. Indeed, as shown in the supplementary Fig. 8Sup the I-V characteristic with 330 meV gap can be directly observed at surface "hot spots". In Fig. 5d, the largest $E_g/E_G \approx 0.7$ indicating that spin ordering may extend beyond visible boundaries of the bubbles as mesoscopic domains [15, 35-38] with $\theta^* \approx 45°$ (or 135°). For pristine MnBi$_2$Te$_4$ a similar tunneling I-V hysteresis was not observed within ±1V bias window; and the gap measured at defect-free areas for MnBi$_2$Te$_4$ was found to be smaller ≈200 meV (see Fig. 9Sup), closely matching Ref. [1] prediction. A larger surface exchange gap in MnBi$_2$Te$_3$ is consistent with decreased depth of Mn atoms after Te layer was removed.

**Discussion**

The overall data analysis including comparison between MnBi$_2$Te$_3$ and MnBi$_2$Te$_4$ shows that (a) large and spatially varying exchange gap E$_g$ has true surface origin and strongly depends on the depth of Mn atoms; (b) the 2D band structure of MnBi$_2$Te$_3$ likely exhibits large effective mass ratio $m_h/m_e \gg 1$, a common feature of nontrivial surface states formed by inverted Bi and Te p$_z$ bands [10] and known for both Bi$_2$Te$_3$ and MnBi$_2$Te$_4$ [1, 10, 34]; (c) spin states in MnBi$_2$Te$_3$ already at RT show the signature of 2D ordering and are tunable by external electric field. The change of $M_\perp$ in the electric field of STM tip can be estimated from our experimental data as

$$\Delta M_\perp \approx 5\mu_B(\delta/E_G) \tag{1}$$

where $5\mu_B$ is the magnetic moment of Mn atoms in MnBi$_2$Te$_4$ lattice [7] ($\mu_B$ – Bohr magneton). In our experiments, the largest field-induced $\Delta M_\perp$ is 2.8$\mu_B$ (per unit cell) in external electric field $10^9$ V/m. The magnetoelectric cross-induction due to coupling of magnetic states and TSS has been predicted for axion insulators [9, 10, 39], with magnetoelectric coefficient $N\alpha/4\pi$ in Gaussian units, where α is the fine structure constant and $N$ is odd number [10]. As we show in the Supplement (P8), at $10^9$ V/m electric field this theoretically predicted response corresponds to $\Delta M_\perp \sim \mu_B$ per unit cell, i.e. very close to what we observed. Moreover, the physical mechanism of cross-induction in our experiment is also related to the coupling between magnetic states and TSS; thus, the observed field-induced switching of spins in MnBi$_2$Te$_3$ likely represents an atomic-scale analogue of quantum topological magnetoelectric effect (TME) in axion insulator. The theoretical description of TME so far has been based on Maxwell's equations [10, 11, 39], whereas its possible atomic scale manifestations have not been discussed prior to our work. We anticipate that further experimental and theoretical exploration of topological quantum metamaterials will bring in deeper understanding of axionic functionalities and their room temperature applications in non-linear optics, spintronics, and quantum information science.

**Methods**

The high-quality single crystals of MnBi$_2$Te$_4$ with $\approx 2 \times 2$ mm$^2$ size have been synthesized for our study using melt-growth method described in the earlier publication [15]. The samples demonstrated n-type electrical conductivity due to unintentional self-doping by vacancies [15]. STM measurements were performed at base pressure $7\times10^{-11}$ Torr at room temperature. Before making ultra-high vacuum (UHV) STM measurements, the crystals were cleaved under UHV conditions using adhesive Kapton tape. The RT cleavage process is most likely responsible for formation of nanocorrals whose presence in as grown MnBi$_2$Te$_4$ samples has not been established [15]. For STM measurements, we used commercial Pt-Ir tips (Bruker Nano Inc.) that were *in situ* cleaned using electron beam heating technique. The tunneling current was measured by preamplifier with signal range ±25 nA. The empirically established in-situ Te layer removal procedure involves scanning MnBi$_2$Te$_4$ at tunneling resistance ~ 100 MΩ during several minutes, typically until detecting significant current noise produced by exfoliated Te flakes. After this procedure, multiple cycles of field-emission cleaning of STM tip are required (see Supplement

P2). The in-situ removal of Te layer occurs at relatively high tunneling resistance compared to other materials [21]; this may be attributed to combination of two factors (a) metastability of $MnBi_2Te_4$ [40] and (b) large transient lateral force induced by STM tip when passing the edges of nanocorrals. All STM images of Te-terminated surface in the manuscript and in the supplement were obtained at tunneling resistances 1.4 GΩ or higher.

## Data availability

The data that supports the plots within this paper and other findings of this study are available from the corresponding author upon request.

**Acknowledgements**

We acknowledge the support for this research from The Ohio State University Center for Emergent Materials (NSF cooperative agreement DMR-1420451) and The Pennsylvania State University Two-Dimensional Crystal Consortium–Materials Innovation Platform (NSF cooperative agreement DMR-1539916).

**Contributions**

I.A. performed most of measurements and prepared Figures 1-5. R.C.W. contributed to STM measurements. S.H.L. and Z.M. synthesized MnBi$_2$Te$_4$ samples. I.A. and Z.M. wrote the manuscript. All authors contributed to discussion of results.

**Competing interests**

The authors declare no competing financial interests.

## Supplementary information

PDF file.

Supplementary Notes P1-P8 and Supplementary Figures 1Sup-9Sup.

**Figure captions**

**Figure 1**. **Atomic structure of MnBi$_2$Te$_4$ and MnBi$_2$Te$_3$.** **(a)** Septuple layers of MnBi$_2$Te$_4$. **(b)** MnBi$_2$Te$_3$/MnBi$_2$Te$_4$ heterostructure obtained by removing top layer of Te atoms using STM tip. The magnetization of Mn atoms is shown by vertical arrows.

**Figure 2**. **Large scale STM images of MnBi$_2$Te$_4$ and MnBi$_2$Te$_3$.** **(a)** The 305×305 Å$^2$ STM image of MnBi$_2$Te$_4$ crystal with Te-terminated surface (sample bias V$_S$=500 mV, I=0.36 nA). **(Inset)** 2D Fourier transform of image in Fig. a reveals 4.3 Å unit cell vectors. **(b)** Cross-section of STM image from Fig. a indicates that the depth of nanocorrals is 6.1 Å. **(c)** STM image obtained after removing top atomic layer (V$_S$=500 mV, I=0.04 nA). The image sizes are 400×400 Å$^2$. The nanocorrals with typical lateral sizes as in Fig. a still can be observed. **(Inset)** The 1700×1700 Å$^2$ STM image obtained after removing terminating Te layer (V$_S$=-256 mV, I=0.08 nA). The atomic layer was removed at ≈500×500 Å$^2$ area indicated by dotted square. The image uses gradient contrast. **(d)** Cross-sections of nanocorrals from Fig. c show that their depth decreased by ≈35%.

**Figure 3**. **High-resolution STM images of MnBi$_2$Te$_4$ and MnBi$_2$Te$_3$.** **(a)** The 130×65 Å$^2$ high resolution STM image of the exposed Bi-terminated surface area (V$_S$=-200 mV, I=0.07 nA). Some of the atoms look brighter than others. These are substitution defects in the Bi atomic plane. **(Inset)** STM image of Te-terminated surface (image sizes 100×30 Å$^2$) shows triplet electronic defects possibly associated with antisites in the Bi atomic plane. **(b)** The 80×150 Å$^2$ high-resolution STM image shows atomic lattices both inside nanocorrals on Mn-termination and outside on Bi-termination (V$_S$=-200 mV, I=0.1 nA). The image uses gradient contrast. **(c, d)**

STM images of subsurface magnetic bubbles ($V_S$=-200 mV, I=0.1 nA). **(e)** STM cross-section from Fig. c. The patch of substitution defects (left side) shows sharp steps; the bubble profile (right side) is smooth and stepless with peak height ≈2 Å due to local DOS variations.

**Figure 4**. **Hysteresis on tunneling I-V characteristics of MnBi$_2$Te$_3$. (a) (Upper inset)** Tunnel I-V characteristics of Bi-terminated surface reveal "counterclockwise" hysteresis loops. **(Main panel)** The same I-V characteristics after applying linear-logarithmic vertical scale arsinh(*I*) [23]. The bandgap E$_g$ can be resolved below the FL. The vertical scale is dimensionless (not shown). The horizontal bar corresponds to I=0. **(Lower inset)** Bubbles of ≈3 nm size can be observed all over the surface and sometimes form clusters ($V_S$=-295 mV, I=0.07 nA). **(b)** Before metal-to-insulator transition, i.e. $V < V_{on}$ (vertical arrow in Fig. a), applying positive tunnel bias leads to tilting of the tunneling barrier with $ln(I) \propto V$; the FL remains pinned at the CBM. **(c)** After metal-to-insulator transition ($V > V_{on}$) the FL unpins from the CBM and shifts inside the gap (band-bending indicated by parallel doted lines). The barrier tilting stops resulting in $I(V) \approx$ const. The E$_g$ increases due to interaction of tip-induced electrostatic field with subsurface Mn atoms, and the increased gap causes hysteresis on reverse I-V sweep. The tip works here as both tunneling sensor and top gate.

**Figure 5**. **Spatial dependence of tunneling I-V characteristics of MnBi$_2$Te$_3$. (a)** STM image of subsurface magnetic bubble indicates locations of I-V spectra measurements in Fig. d. **(b, c)** The electric field of STM tip and the suggested spin structure of the bubble are shown before (b) and after (c) metal-to-insulator transition of TSS. In Fig. b the residual spin canting outside the bubble is not shown. The magnetic defect is indicated by a dotted circle. **(d)** The linearized

tunneling I-V spectra obtained at the same tip-sample distance at five different locations (white dots in Fig. 5a). The correlated spatial variations of $E_g$ and $\delta$ can be observed (including point 1A not shown here where I-V spectrum was measured at different *R*; see Fig. 6Sup). **(e)** Simulated dependences of $\delta/E_G$ and $E_g/E_G$ on the distance from the center of the bubble ($E_G$ - gap for zero canting angle $\theta$). Dashed and dotted lines correspond to $w_1$=5 nm (zero-field value) and $w_2$=1 nm (in-field value) respectively. The solid line shows the difference between these two curves quantifying the size of the I-V hysteresis loop.

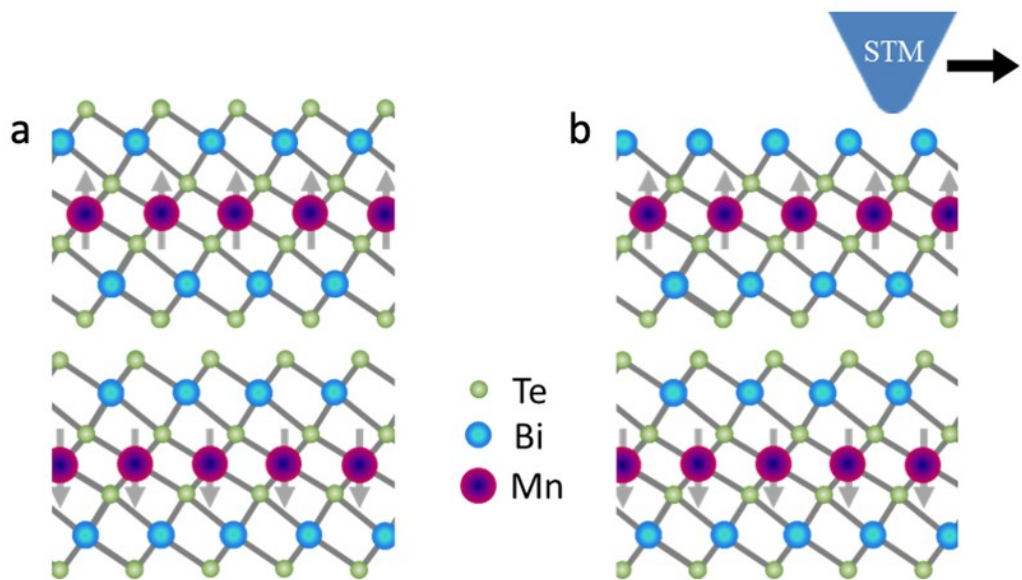

Altfeder et al. Figure 1

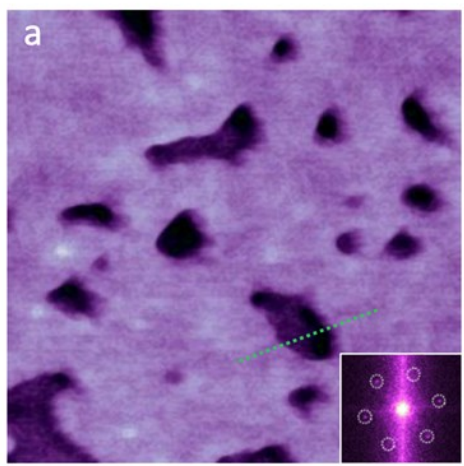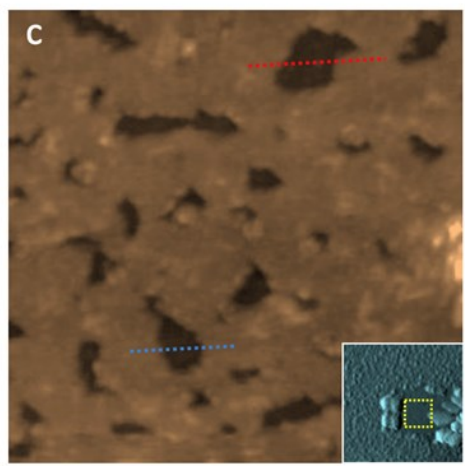
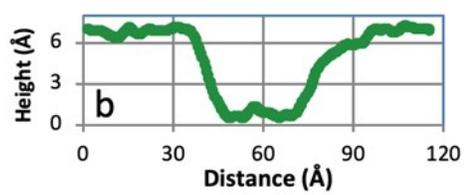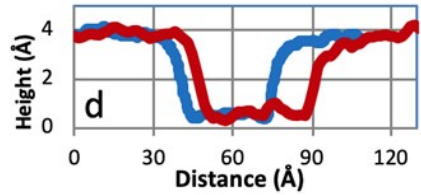

Altfeder et al. Figure 2

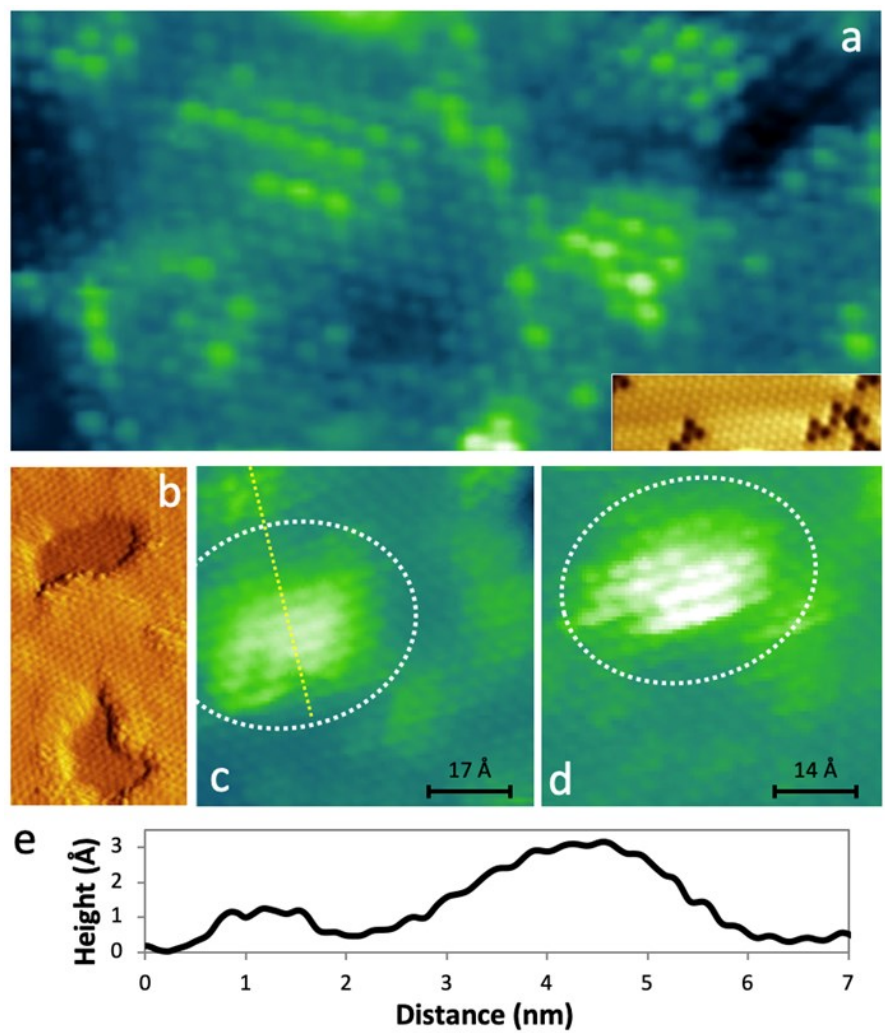

Altfeder et al. Figure 3

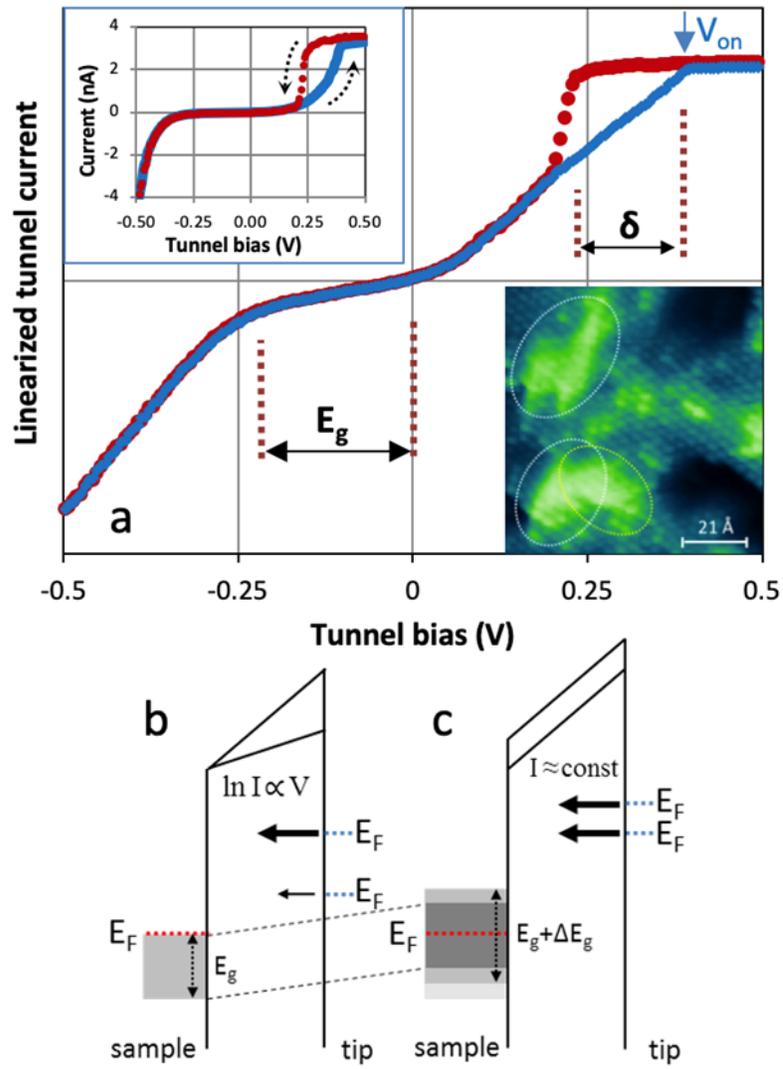

Altfeder et al. Figure 4

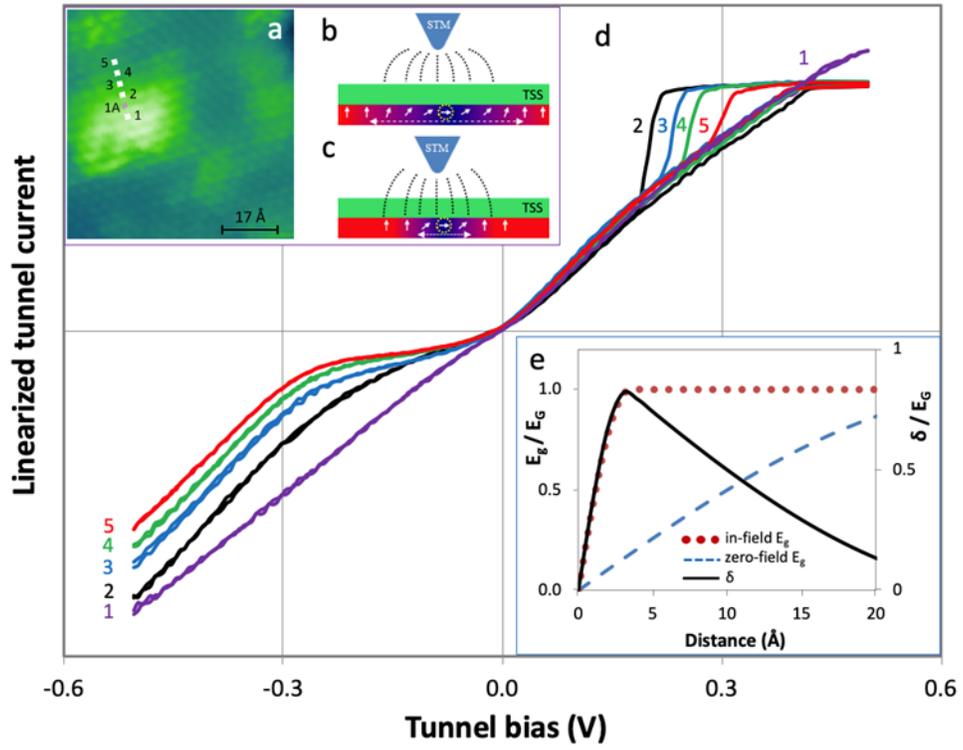

Altfeder et al. Figure 5